\documentclass[twocolumn,showpacs,floatfix,prl]{revtex4}
\usepackage{graphicx}% Include figure files
\usepackage{dcolumn}% Align table columns on decimal point
\usepackage{bm}% bold math
\def\be{\begin{equation}} 
\def\ee{\end{equation}}
\def\bea{\begin{eqnarray}} 
\def\eea{\end{eqnarray}}
\def \line{\hbox to \hsize}    
\def\frac #1#2{{#1\over #2}}

\def \ket #1{{\vert #1\rangle}}
\def \bra #1{{\langle #1\vert}}
\def \brak #1#2{{\langle#1\vert#2\rangle}}

\def\1{\mbox{\bf 1}}
\newcommand{\comment}[1]{}

\begin{document}
%\draft %(only for revtex) 

\title{Three dimensional topological invariants for time reversal invariant Hamiltonians and the three dimensional quantum spin Hall effect}
\author{Rahul Roy}
\affiliation{University of Illinois, Department of Physics\\ 1110 W. Green St.\\
Urbana, IL 61801 USA\\E-mail: rahulroy@uiuc.edu}

\begin{abstract}
  The $Z_2$ invariant for filled bands in the ground states of systems  with time reversal invariance characterizes the number of stable pairs of edge states. Here we study the $Z_2 $ invariant using band touching methods discussed in a recent previous work \cite{roy2006zcq}
 and extend the study to three dimensions. Band collisions preserve the $Z_2 $ invariant both in two and three dimensions, but there are crucial differences in the two cases. In three dimensions, we find a novel fourth $Z_2 $ invariant which is characterized by a ``trapped monopole" in momentum space. If the monopole charge in half the Brillouin zone is odd, then atleast one of the monopoles cannot recombine with another monopole and vanish unlike the case when the monopole charge is even. We also point out the possibility of a three dimensional quantum spin Hall effect and discuss the connection of various topological invariants to such an effect.            
\end{abstract}

\maketitle

\section{Introduction}

   A new invariant that characterizes ground states of two dimensional systems in a periodic potential with time reversal invariance when the Fermi Energy lies in a gap was introduced recently by Kane and Mele  \cite{kane2005zto}. They discovered this invariant while studying a model for graphene which exhibits the quantum spin Hall effect \cite{kane2005qsh}. The model has a single pair of edge states which are robust to small perturbations. The robust edge states are also seen in an earlier model for the quantum spin Hall effect based on Landau levels in the presence of a strain gradient \cite{bernevig-2005-} and a model for the square lattice \cite{roy2006iqh}.  
    The robustness of an odd number of  time reversed pairs of edge states can be explained  on the basis of the inability of a state to scatter with its time reversed counterpart \cite{xu-2006-73,kane2005zto,wu2006hla}. A formula for the invariant which involves counting the number of zeroes of the Pfaffian of the time reversal opearator among bands  was provided in \cite{kane2005zto}. This characterization, however, does not establish a link between the topology in momentum space and the robustness of the edge states  as observed in numerical studies \cite{sheng-2005-95}. In a previous paper \cite{roy2006zcq}, we showed that the $Z_2$ invariant could be constructed as an obstruction in momentum space. The $Z_2$ invariant was also cast in terms of the more familiar Chern numbers thus leading to a connection between the robust edge states and the bulk topology, similar to the one in the integer quantum Hall effect.  Using the additivity of Chern numbers, the $ Z_2$ index for the multi-band case was also written down. 

    In this paper, we extend the study of $ Z_2$ invariants to three dimensions. We begin with a review of the explanation of the invariance of the $ Z_2$ index in two dimension elaborating the band touching arguments presented in \cite{roy2006zcq}. We then extend these methods to study the  $Z_2$ invariants in three dimensions.  We find a total of four independent $Z_2$ invariants, one for each of the faces of the Brillouin zone torus and a fourth one which characterizes the monopole charge in momentum space. A nontrivial value of this index implies the existence of  a set of ``trapped monopoles" in momentum space which unlike the ordinary case, cannot recombine and are hence stable. The monopole charge of an exotic insulator with a non trivial fourth $Z_2$ index could however change in the presence of a magnetic field paving the way for a possible experimental observation.  
    We discuss the possibility of a three dimensional counterpart of the recently proposed spin quantum Hall effect \cite{bernevig2006qsh}. We point out that while the stability of the edge states in such a model is a consequence of the $ Z_2 $ invariants, the stability of the quantized response of the spin Hall conductivity is described by a set of three  topological invariants in the space of twisted boundary conditions, namely, the spin Chern numbers \cite{sheng-2006-}, which were recently introduced to study the stability of the quantized spin Hall response in two dimensions.

  \section{Topological $Z_2$ invariants}
  \subsection{$Z_2$ invariants in two dimensions}    
 We begin with a study of the $Z_2$ invariant for a ground state consisting of  time reversed sets of bands in two dimensions. First, consider the simplest case when we have a single pair of time reversal invariant bands in the ground state. 
 Consider a case when the bands do not have degeneracies at points other than the ones at which such a degeneracy is required by time reversal invariance. Note that at degeneracies required by symmetries, in general, the bands can still be identified. 
   The Hamiltonian in this case can be written as 
  \bea
   H = H_1 + H_2
  \eea
 where
 \begin{eqnarray*}
  H_1 &=& \int dk \tilde{H_1} (k)\\
  & = & \int dk  \left[E_1(k) \ket{u_{I}(k)}\bra{u_{I}(k)}  + E_2 (k) \ket{u_{II}(k)} \bra{u_{II}(k)} \right] \\
  H_2 & = & \int dk \sum_{n, E_n (k) >0 } E_{n}(k) \ket{n(k)}\bra{n(k)} 
 \end{eqnarray*}
  in which $ k = (k_1, k_2), dk = dk_1 dk_2 $ and $ E_1, E_2 < 0 $, and the Fermi energy, $E_f = 0$. 
   The time reversal invariance of the Hamiltonian implies that 
   \bea
    E_1(k) = E_2(-k)
    \eea
    where $ (-k) = (-k_1, -k_2) $
    and the bands $ u_{I}(k), u_{II}(k)$ can always be chosen such that
    \bea
     \Theta(\ket{u_{I}(k)}) = \ket{u_{II}(-k)}, \Theta\ket{u_{II}(k)}= -\ket{u_{I}(-k)} 
    \eea
    where $ \Theta $ is the time reversal operator. 
    Now imagine that we adiabatically change the Hamlitonian $H(t)$ as a function of some variable t. As long as the ground state is adiabaticaly connected to that of the original Hamiltonian and the bands do not touch, the Hamiltonian can always be written as  above for some suitably redefined bands. The Chern numbers for the bands are then individually conserved. The time reversed pair of bands have a total Chern number of zero. (This follows from homotopy arguments of the form presented in \cite{roy2006zcq} or by a direct computation of the Chern number in terms of the Berry phase such as by using the TKNN formulae \cite{PhysRevLett.49.405}. A simple physical argument for this follows from the observation that the Hall conductivity of a system with time reversal invariance is zero. One can always move the bands so that they are the only ones that contribute to the Hall conductivity. Then using the standard result \cite{PhysRevLett.49.405}, it follows that the sum of the Chern numbers of these bands is zero.)
    
     Now suppose that a degeneracy  occurs at a point $( t_0 , k_0)  $.  Due to time reversal invariance, the bands must also touch at $ (t_0, -k_0) $.  It is known that a two level degeneracy is generic while a three or higher level degeneracy is not. Further, non-generic three level degeneracies can be adiabatically continued to sets of two level degeneracies. Without loss of generality therefore,  one can always write $\tilde{H_1 }(t)$  in the vicinity of the degeneracy points as
    \bea
    \tilde{H}_1 (t,k) = \vec{m}(t,k).\vec{\sigma} + {E_I + E_{II} \over 2} I
    \eea
    where the two indices of the $\sigma $ matrices correspond to the two bands $\ket{u_{I} (k)} $ and $ \ket{u_{II}(k)} $. 
  Time reversal invariance then dictates that
   \be 
    \vec{m}(t,-k) = - \vec{m}(t,k) 
    \ee 
    At the degeneracy point, $ \vec{m}(t_0,k_0) = \{0,0,0\} $.
    
    As the parameter t is further changed, the bands may split apart again. However, as they separate, the bands might now have a new set of Chern numbers whose sum has to be the same as  before the band collision(zero) \cite{PhysRevLett.51.51}. One can visualise the degeneracy points as monopoles which flow in and out of the bands. 
    
    If one sets $ \omega = \vec{m}/|m| $, then the Chern number exchange between the bands is given by \cite{bellissard-1995-}
   \bea
     n(x_0) = \pm { 1 \over 4 \pi} \int_{\Sigma_j} \brak{\vec{\omega}}{d\vec{\omega}\wedge d\vec{\omega}} 
   \eea
   where $ \Sigma_j $ is the surface of a small sphere enclosing the degeneracy point in the three dimensional space of points $(t, k_x, k_y) $. 
    From the property of time reversal invariance, it follows  that the Chern number exchange between a band and its time reversed counterpart at the two points $(t_0, k_0 ) $ and $ (t_0, -k_0) $ is exactly the same (in magnitude and sign). Hence the total Chern number exchanged between bands is always an even number. 
    Next  consider the effects of band collisions for ground states consisting of multiple pairs of bands. Two possible kinds of collisions can occur, the first, a degeneracy between a band and its time reversed counterpart, and the second, a degeneracy between two bands which are not related by time reversal. In the second case, however, the fact that the time reversed pairs before and after collision have opposite Chern  numbers which add up to zero for each pair implies that if a band $ \ket{\alpha}$ changes its Chern number by  $ n_{\alpha}$ then the Chern number of $ \Theta \ket{\alpha}$ changes by 
    $ -n_{\alpha}$. These preserve the net $Z_2$ invariant presented in \cite{roy2006zcq}
    \bea
    E = |\sum_{c_n >0} c_n | \textrm{mod} 2
    \eea      
    The first kind of process also obviously preserves this invariant from the previous arguments. Thus we have demonstrated that there is an index, E which stays invariant when the Hamiltonian is changed in a continuous way such that the Fermi Energy always lies in a gap. 
    
     \subsection{$Z_2$ invariants in 3 dimensions}
     Let us parametrize the three torus, $T^3 $ by $\{-1\geq x,y,z \leq 1 \}$. In three dimensions, the time reversal operator, $ \Theta $ maps the fiber at the point 
     $(x,y,z)$ to the point $ (-x,-y,-z)$. Let  $ A $ be the map on $T^3 $ induced by the map $\Theta $ on the fiber bundle of the bands. The planes $ x,y,z = 0, 1 $ get mapped to themselves and therefore one can associate $Z_2 $ invariants with each of them. The planes $ x= y, \, x=-y,\,  x=z,\, x=-z,\,  y=z,\,  y=-z $ are an additional set of planes, which get mapped to themselves.
We consider separately two different cases :

\begin{enumerate}
\item The bands do not touch at any points in $T^3$. There are then three independent Chern numbers for each band and the Chern number of the ground state is again zero. 
       We can define a set of three $Z_2 $ invariants $E_{yz}, E_{zx},E_{xy} $ for each of the planes $ x=0, \, y=0 $ and $z=0 $ using the formula for the two dimensional case. These are identical to the $Z_2$ invariants of the planes $ x= 1$ etc.
        Thus    $ E_{xy} = | \sum_{c_{n} ^{xy} > 0 } c_{n} ^{xy}| \,\textrm{mod} \,2 $ where $c_{n} ^{xy}$ is the Chern number of the nth band in the $xy$ plane and similarly for $ E_{zx}$ and $ E_{xy}$. 
 Now, the $ Z_2 $ invariants for the planes $ x=y $ etc. can be obtained as sums of the $Z_2 $ invariants of the $E_{xy} $ etc, since the Chern numbers for these planes can be obtained as sums of the Chern numbers of the planes $ x=0, \, y=0 $ and $z=0 $ . Let us now perturb $H_1(x,y,z,t)$ as before. Degeneracies can occur and split into Dirac points. These can be thought of as monopoles which can recombine after relative displacement through a reciprocal lattice vector 
 \cite{haldane-2004-93}.  
 The Chern numbers for the planes are independent and can be changed independently. At the same time, because of time reversal invariance, the Chern numbers can  only change in such a way that the $Z_2$ invariants are preserved in the process. There are thus only three independent $Z_2$ invariants and the $Z_2$ invariants for the other planes depend on these.  Thus the set of all Ground states with TR invariant Hamiltonians with no bands crossing the Fermi state and no band touching in $T^3$ can be classified by $ Z_2 ^3 $. 

\item
   Now let us consider ground states where the bands do touch. 
   Let us call the manifolds
  \bea
   T^3 _{{1\over 2} z+} = [ 0\geq z \leq 1,  -1\geq x \leq 1,  -1\geq y \leq 1] 
  \eea
  \bea
  T^3 _{{1\over 2} z-} = [ -1\geq z \leq 0,  -1\geq x \leq 1,  -1\geq y \leq 1]
  \eea
  and similarly for the x and y directions,  3-d half tori. We further divide these half tori into quarter tori which are defined by the intersection of half tori, such as 
  $ T^3_{{1\over 4} z-,x-} = T^3 _{{1\over 2} z-} \bigcap T^3 _{{1\over 2} x-} $. 
   The half and quarter tori come in pairs such as $\{T^3 _{{1\over 2} z\pm}\}$ which get mapped onto each other under the projection of the time reversal operator. We call these ``complementary pairs". 
     For simplicity, we consider at first a single pair of time reversed bands and a set of generic Dirac like degeneracy points which can be thought of as monopoles, which do not lie at the surfaces of any of the quarter tori,  though the arguments can be easily modified to adapt to non-generic degeneracies and more complicated cases. Time reversal maps these monopoles in quarter tori to monopoles in  their complementary quarter tori such that if a band collides with its time reversed counterpart, the charge that flows in at one monople to a band is the opposite of the charge that flows in at the other monopole. If there is an odd number of such monopoles in the half torus $ T^3 _{{1\over 2} z-}$, then the $Z_2$ indices of the plane $ z= 0 $ and  the plane $z=1 $ are different, while they are the same if there are an even number of monopoles. Since these $Z_2 $ indices are homotopically stable, atleast one pair of monopoles in the Brillouin zone cannot recombine and vanish in the first case. These monopoles are thus trapped and hence stable. Further, from the mapping of monopoles to their counterparts in quarter tori, one can see that if a certain half torus has an odd number of monopoles, so must all the other half tori. Hence there are no additional $Z_2$ invariants. In the case of multiple bands, one may simply evaluate the monopole charge flowing into half the set of filled bands, say for instance the set of bands with positive Chern numbers, and half the set of those with Chern number zero. There are thus four independent $ Z_2 $ invariants characterizing time reversal independent  ground states with no bands crossing the fermi surface. An analysis through K theory seems to suggest that the fourth $Z_2$ invariant or the monopole charge described above, arises from the
twisted KR theory of $S^3$ while the remaining three arise from the
faces of $T^3 $ \cite{work-in-progress}.
    When time reversal symmetry is broken, such as in the presence of an external magnetic field, the $Z_2 $ indices are no longer invariant, and hence the trapped monopoles may vanish. This might lead to an experimental observation of such trapped monopoles. 
\end{enumerate}
  
  \section{Quantum spin Hall effect in three dimensions}  
 Recently it has been suggested that the spin hall conductivity can also be quantized in two dimensions \cite{bernevig-2005-} and various models in which such a quantization occurs have since been proposed \cite{ kane2005qsh, qi-2005-, roy2006iqh}. Here we show that a quantized spin Hall response is also possible in three dimensions and discuss the relevance of various topological invariants to such an effect.
    It is well known that in two dimensional electron systems with a periodic potential, the Hall conductivity in the presence of an external magnetic field is quantized to an integer known as the Chern number when the Fermi energy lies in a gap of the extended states \cite{PhysRevLett.49.405}. The three independent Chern numbers for filled bands in three dimensions play the same role in determining the quantized Hall response as the single Chern number does in two dimensions \cite{Halperin-3dqhe,kohmoto1992det, deblas2004det}. 
        The quantized spin Hall  current in two dimensions arises from equal contributions from the up and down spins, $ {\bf J_s }= {\hbar \over 2e}( {\bf J_{\uparrow} - J_{\downarrow}} ) $, each of which has a quantized response to the external Electric field, but in opposite directions.  Thus one expects that in three dimensions, when the Fermi energy lies in a gap, the spin conductivity tensor can be written in the form : 
 \bea
  \sigma^{s} _{ij} =  {e \over 8\pi^2 h} \epsilon_{ijk}G_k
  \eea  
  where $\vec{G}$ is a reciprocal lattice vector. 
   When a well defined Fermi surface exists, the non quantized portion of the spin conductivity tensor can be written in terms of an integral over the Fermi surface \cite{haldane-2004-93}.
       A Hamiltonian which displays such a quantized response  in the $ k_z= 0$ and $ k_y =0$ planes can be written based on  the two dimensional model presented in \cite{roy2006iqh}.  
     The  stability of the edge states is governed by the three dimensional $Z_2 $ invariants presented earlier, even though the spin Hall conductivity itself is not expected to be quantized when terms which mix up and down spins are present. An explicit formulation of the edge states for the graphene model was presented in \cite{sengupta-2006-} and a general connection between bulk topology and edge states was discussed in \cite{qi-2005-}.
     Since a non trivial value of the fourth $Z_2 $ index is not possible without band degeneracies that change the $Z_2 $ invariants of a pair of bands across a half torus, a quantized spin Hall response can only be obtained when this index is zero. 
     
     One can instead define three spin Chern numbers following \cite{sheng-2006-} for the many body wavefunction of the ground state:  
\bea
C^{\alpha,\beta} _i   = \epsilon_{ijk} {i \over 4\pi} \int \int d\theta^{\alpha}_{j}d\theta^{\beta}_{k} \brak{{\partial \psi \over \partial \theta^\alpha _j}}
{{\partial \psi \over \partial \theta^\beta _k}}
\eea
where $ \epsilon_{ijk}$ is the three dimensional antisymmetric tensor where $ \alpha = \{\uparrow, \downarrow\}$ and the variables $\theta^{\alpha} _j $ represent the twisted boundary conditions following the notation of \cite{sheng-2006-}. These are a more appropriate measure of the robustness  of the spin hall conductivity to the presence of spin independent disorder. \\

   The $Z_2 $ invariant is also discussed in a few other very recent preprints. In \cite{fu-2006-}, the role of this invariant in a one dimensional quantum spin pump is discussed, and the role of the index as an obstruction was presented and in \cite{fukui-2006-taqshe}, the connection between the spin Chern number and the $Z_2$ invariant was discussed. \\

 While this work on 3-d invariants and the 3-d quantum spin hall effect  was in progress, a preprint \cite{moore-2006-} appeared which discusses the 2-d and 3-d invariants using results from homotopy theory. Our results seem to be in agreement with the ones obtained in that work using different arguments. 

  I would like to gratefully acknowledge extensive discussions with Prof. Sheldon Katz and Prof. Michael Stone which have greatly helped shape my understanding of the subject matter and thank them for the same. I also thank the Physics department at UIUC for support.
\bibliography{qshe}
\end{document}